\documentclass[cits]{PoS}
\newcommand{\met}{\hbox{E\kern-0.5em\lower-0.1ex\hbox{/}}_T}
\newcommand\simlt{\lower.5ex\hbox{$\; \buildrel < \over \sim \;$}}
\newcommand\simgt{\lower.5ex\hbox{$\; \buildrel > \over \sim \;$}}

\title{Relativistic Flows in TeV Blazars}

\ShortTitle{Relativistic Flows in TeV Blazars}

\author{\speaker{Amir Levinson}\thanks{This work was supported by an ISF grant for the Israeli Center for High Energy Astrophysics.}\\
        Raymond and Beverly Sackler School of Physics and Astronomy, Tel Aviv university, Tel Aviv 69978, Israel\\
        E-mail: \email{Levinson@wise.tau.ac.il}}

\abstract{Rapid variability of the TeV emission in several blazars implies a central black hole mass 
$M_{BH}<10^8M_\odot$, appreciably smaller than the values estimated from the $M_{BH}-L_{bulg}$ relation, and 
Doppler factors for the $\gamma$-ray emitting fluid much larger than those associated with 
radio patterns.  We discuss the conditions in the central engine required to account for the short timescales 
and large luminosities observed, and propose some explanations for the inferred kinematics of the source on 
various scales.}

\FullConference{Workshop on Blazar Variability across the Electromagnetic Spectrum\\
		 April 22-25 2008\\
		 Palaiseau, France}

\begin{document}
\section{Introduction}
Recent observations of TeV AGNs raise new questions regarding the parameters of the central engine, and the location and 
kinematics of the TeV emission zone.   In particular, the rapid flares reported for Mrk 501 and PKS 2155-304
appear to indicate a relatively small black hole mass, which is difficult to reconcile with the black-hole/bulge
relation, and accretion rates much larger than previously thought.  The systematic differences between the Doppler 
factor inferred from TeV observations and that associated with radio knots, and the rapid variations of the 
X-ray flux emitted from the HST-1 knot in M87, a region located at a distance of about $10^6 (2GM/c^2)$ from the central  
black hole, are also issues of interest and may be related.
These observations motivate reconsideration of the standard view, according to which 
dissipation of the bulk energy on sub-parsec scales is accomplished
predominantly through formation of internal shocks in colliding fluid shells.  Below we review 
these issues in greater detail and speculate about possible explanations for some of the open questions. 

\section{Conditions in the central engine}
The characteristic size of any disturbance produced by a central engine having a radius $r_g$ must satisfy $d\simgt r_g$
in the rest frame of the engine.  As a consequence, the associated variability time, if produced by temporal
fluctuations of the ejected fluid, cannot be much shorter than $r_g/c$, regardless of Doppler boosting.
The variability may imprint the scale of some external disturbance, e.g., in case of a collision of
a fluid shell expelled from the central source with some intervening cloud of size $d<<r_g$.  However, in that case
only a fraction $\sim (d/r_g)^2$ of the bulk power released by the engine can be tapped to produce the
observed radiation.  If indeed reflects the size scale of the engine, a variability time
$t_{\rm var}=300t_{300}$ sec, with $t_{300}\simeq 1$ reported for PKS 2155-304, would imply 
a black hole mass of $M_{BH}\simlt5\times10^7 M_{\odot}t_{300}$.  The isotropic equivalent luminosities
measured for the rapid TeV flares, $L_{TeV}$, are quite large; for the bright flares observed in 2155-304 
$L_{TeV}\simgt10^{46}$ erg s$^{-1}$.  For a two-sided conical jet with an opening angle $\theta=0.1\theta_{-1}$  , 
the temporary jet power released during the TeV flare must satisfy 
\begin{equation}
L_j>f_bL_{TeV}\simeq10^{44}\theta_{-1}^2L_{TeV,46} \qquad {\rm erg\ s^{-1}},\label{L_j} 
\end{equation}
with $f_b=\theta^2/2$ denoting the beaming factor of the emission.  A plausible powering mode is
the Blandford-Znajek (BZ) mechanism.  The BZ power that can, in principle, be 
extracted from a rapidly rotating black hole can be expressed as 
\begin{equation}
L_{BZ}=10^{45}\epsilon B_4^2M_8^2 \qquad {\rm erg\ s^{-1}},\label{LBZ} 
\end{equation}
where  $M_8=M_{BH}/10^8M_{\odot}$ is the black hole mass in fiducial units and  $B=10^4B_4$ G is the strength of the
magnetic field threading the horizon.  The efficiency factor $\epsilon$ depends on the geometry of the magnetic 
field and other details, and typically $\epsilon\simlt 0.1$.  Taking $L_j=L_{BZ}$ and using Eqs. (\ref{L_j}) 
and (\ref{LBZ}) we find that a field strength of
\begin{equation}
B_4\simgt2(\epsilon/0.1)^{-1/2}t_{300}^{-1}\theta_{-1}L^{1/2}_{TeV,46} \label{B}
\end{equation}
is required to account for the rapid TeV flares observed.  Note that the last equation holds also in 
situations where the flare results from a collision of the jet with a
cloud, provided that a fraction $(d/r_g)^2$ of the bulk power can be converted to radiation. 
In that case Eq. (\ref{L_j}) should be modified to $L_j>(r_g/d)^2f_bL_{TeV}$, with $d=ct_{\rm var}$, so that 
equating $L_j$ and $L_{BZ}$ yields essentially the same result. 
Taking the energy density of the magnetic field in the disk to be a fraction $\eta$ of equipartition yields
\begin{equation}
B_4\simeq 5\sqrt{\eta\dot{m}\over M_8} \label{Beq}
\end{equation}
for the field in the vicinity of the horizon, where $\dot{m}$ is the accretion rate in units of Eddington.
For $\eta\sim 0.1$ and $M_8=0.5$ accretion rates near the Eddington limit are needed to satisfy Eq. (\ref{B}).
It could be that the rapid flares result from enhanced accretion episodes, in which case a large disk luminosity
is anticipated during the flares.  Whether this is consistent with observations and models for the TeV emission is yet
to be explored.  Alternatively, the radiation from the disk may be suppressed if the hole accretes in some radiative
inefficient mode. 

The limit on the black hole mass derived above appears to be inconsistent with the black-hole/bulge relation \cite{McL02,woo05}, which for
PKS 2155 yields $M_8\sim20$ \cite{ah07}.  It is plausible that this apparent inconsistency simply reflects 
the scatter around the $M_{BH}-L_{bulg}$ 
relation \footnote{Even though the scatter in a sample consisting of normal galaxies, Syferts and QSOs 
is claimed to be only a factor of 3 roughly  \cite{McL02}, one should be cautious in applying those estimates to 
an individual object of a certain type}.  Alternatively, it could be that the source imprinting the variability has a characteristic size 
considerably smaller than the black hole's horizon as, e.g., in the situation mentioned above whereby the jet collides with an intervening cloud.
In both cases the magnetic field strength required to account for the energetics is given by Eq. (\ref{B}). 
A smaller field is needed if substantial focusing of the outflow, in a manner described
below, occurs on relevant scales.  The observed variability may then be attributed to dissipation in internal shocks or blobs produced by 
reflection at the axis, as explained in \S 4.3, that may have a characteristic scale associated with the cross-sectional radius $a$,
or thorough collision with an obstacle. Whether such focusing can be accomplished close enough to the central source is yet unclear.
Another possibility, proposed in Ref.~\cite{Der08}, is that the bulge in PKS 2155 hosts a binary
system consisting of the black hole producing the TeV jet in orbit around an inactive $10^9 M_\odot$ black hole.

\section{Kinematics}
A lower limit on the Lorentz factor of the emitting fluid can be derived from the requirement that the pair 
production opacity at TeV energies is sufficiently small to allow escape of the TeV photons to infinity.
The size of the emission region (as measured in the Lab frame) implied by large amplitude variations
with a variability time $t_{\rm var}=1 t_{\rm var,h}$ hour is limited to $\Delta r\simlt10^{14}\Gamma D t_{\rm var,h}/(1+z)$ cm, but is expected not to be much smaller than $r_g$, as explained above.  Here $\Gamma$ and $D$ are the bulk Lorentz factor and the corresponding Doppler factor of the emitting matter, respectively, and $z$ is the redshift of the source.   The condition that the absorption length of the observed TeV photons is larger than $\Delta r$ then yields,
under the assumption that the target synchrotron photons originate from the same region,   
a limit on the Doppler factor of the TeV emission zone independent of its location.  Detailed calculations  
suggest $D\sim 30 -100$ for extreme events \cite{kra02,lev06,beg08}.  Such high values appear to be consistent with those obtained from fits of the SED to a homogeneous SSC mode, but are in clear 
disagreement with the much lower values inferred from unification schemes \cite{Urry91,Hardcastle03} and superluminal 
motions on parsec scales \cite{mar99,Giroletti04,Jorstad01}.  

The assumption that the synchrotron emission originates from the same region producing the TeV flares is crucial
for the estimates of the Doppler factor mentioned above.
It has been argued that such high values of the Doppler factor may not be required if the 
$\gamma$-ray production zone is located far from the black hole, at radii $r_{\rm em}>>\Delta r$.
This possibility is motivated by recent observations of M87, as described below.
In that case the compactness of the TeV emission zone may be constrained by the variability of the IR flux observed simultaneously
with the TeV flare, allowing low values of $D$ in cases where the variability time of the IR emission is much longer than
the duration of the TeV flare.  
However, such a mechanism requires either, a jet power much larger than the luminosity of the TeV emission measured 
during the flare, $L_{TeV}\sim10^{46}$ erg s$^{-1}$, or focusing of the jet.

\section{Where is the location of TeV emission zone?}
The location of the $\gamma$-ray emission zone is yet an open issue.  Different emission sites have been 
considered in the literature, including the black hole magnetosphere, the jet base, and intermediate scales,
as speculated in the case of M87.  We discuss these in some detail below.

\subsection{The black hole magnetosphere}
Under certain conditions the TeV emission can, in principle, be produced in the black hole magnetosphere in a 
{\em pulsar-like} process.  This scenario 
is motivated by the original work of Boldt and Ghosh \cite{BG99} who proposed that starved black 
hole magnetospheres in dormant AGNs may provide acceleration sites for the observed ultra-high energy cosmic rays (UHECRs),
and the observation that the accelerated particles should produce detectable curvature TeV emission \cite{lev00}.  
The maximal electric potential difference that can be generated by a rapidly rotating black hole is
\begin{equation}
\Delta V\sim 4.4\times10^{19} B_4M_8 \qquad {\rm volts}. \label{DeltV}
\end{equation} 
For $M_8\simgt10$, $B_4\sim1$ this potential is marginally sufficient to account for the 
highest UHECRs observed.  If vacuum breakdown is prevented, as discussed below, 
then a stray particle entering the gap should be accelerated, 
in the absence of severe energy losses, to energies near the full voltage.
However, as shown by Levinson \cite{lev00} the accelerated particles
suffer severe energy losses, owing to curvature radiation, that limit their energy to values substantially smaller than 
the maximum potential drop given in Eq. (\ref{DeltV}).   The peak energy of the curvature photons
depends on the charge $Z$ of the accelerated particle but not on its mass.  Denoting by $\tilde{\rho}$ the 
curvature radius of magnetic field lines in the gap in units of $r_g$, this energy can be expressed as 
\begin{equation}
\epsilon_{\gamma,m}\simeq 1.5 M_8^{1/2}(B_4/Z)^{3/4}\tilde{\rho}^{1/2}\tilde{h}^{3/4} \qquad  {\rm TeV},
\label{Eg}
\end{equation}
where $\tilde{h}$ is the dimensionless gap height.  The total power released in the form of UHE particles and curvature 
photons depends on the rate at which charged nuclei are injected into the magnetosphere.  To account for the observed 
flux of UHECRs only a small fraction of the total power available, $P=\rho_{GJ}\Delta V cr_g^2$ where 
$\rho_{GJ}$ denotes the Goldreich-Julian electric charge density, is required on the average from a single source,
consistent with the assumption that the parallel electric field in the gap is unscreened (i.e., $n<<n_{GJ}$).  Nonetheless, 
for sources within the GZK sphere the predicted flux of curvature TeV photons is above detection limit of current 
experiments, at least for some.  In Ref. \cite{lev00} 
it has also been speculated that the episodic TeV emission detected in some TeV blazars may also be due to this mechanism.
However, if the black hole mass is as small as inferred above for Mrk 501 and PKS 2155, $M_8\simlt0.5$, it may be difficult
to account for the highest energy photons observed.  Inverse Compton scattering of ambient radiation by electrons/positrons 
accelerated in the gap is a more likely origin of magnetospheric TeV emission.
In the case of M87 the observed spectrum extends up to an energy of about 10 TeV.  Adopting $M_8=30$, $\tilde{\rho}=\tilde{h}=1$
in Eq. (\ref{Eg}) implies $B_4\simgt1$ for a curvature emission origin.   This value is about $10^3$ times larger
than the equipartition field (see Eq. \ref{Beq}) inferred from the Bolometric luminosity, $L\simeq10^{-6}L_{Edd}$.
However, given current estimates of the jet power, $L_j\simgt10^{44}$ erg s$^{-1}$ \cite{bick96,staw06}, it is 
roughly consistent with the field strength required for magnetic launching of the jet (for a BZ scaling 
Eq. \ref{LBZ} with $M_8=30$, $\epsilon\simlt0.1$ implies $B_4>0.05$).  Alternatively, the TeV photons may be 
produced through IC scattering of ambient radiation \cite{Ner07}.  In 
that case an equipartition field, $B_{eq4}\simeq10^{-3}$, is sufficient to account for the observed flux and spectrum.

Breakdown of the gap will occur under conditions that allow formation of intense pair cascades.
The curvature photons will initiate a cascade if the pair multiplicity is sufficiently large.
Detailed analysis \cite{lev00} demonstrates that this is expected when the strength of the magnetic field
threading the horizon satisfies $B_4> 13 M_8^{-2/7} \tilde{\rho}^{2/7}/\tilde{h}^{5/7}$.  Alternatively, 
vacuum breakdown may occur through the agency of ambient radiation.   This in turn requires an external source of photons that 
can i) be Compton scattered to sufficiently high energies by the electrons 
accelerated in the gap, and ii) contribute high enough opacity for pair 
production with the scattered gamma rays.  The latter condition requires a luminosity  
$L>10^{37}M_8 \tilde{R}^2(\epsilon_s/0.1 {\rm eV})$ erg         
s$^{-1}$, where $\tilde{R}$ is the size of the external radiation source in units of $r_g$ and $\epsilon_s$
its peak energy.  The former condition is satisfied if the magnetic field strength 
is large enough to allow electrons to be accelerated to a Lorentz factor
$\gamma>m_ec^2/\epsilon_s$, that is, $B_4>10^{-5}(L/10^{40} 
{\rm erg\ s^{-1}}) (\epsilon_s/0.1 {\rm eV})^{-2}\tilde{R}^{-2} 
M_8^{-1}$.  This condition is probably satisfied by all AGNs.  To avoid vacuum breakdown in TeV blazars 
requires accretion channel with low radiative efficiency (note that this automatically guarantee
that TeV photons can escape).  One might worry about other opacity sources that can absorb the TeV
photons emitted from the magnetosphere, in particular the synchrotron
photons produced in the jet.  The observed 
synchrotron flux in this scenario should be produced at sufficiently large radii to allow escape of the TeV photons.

\subsection{The inner jet}
A natural expectation, until recently, has been that the $\gamma$-ray
emission is produced at small radii, near the base of the jet.  Naively, it is anticipated that disturbances will
dissipate bulk energy on scales $r_{diss} \simgt \Gamma^2 r_g$.  For TeV blazars $r_{diss}\simgt10^{16}$ cm or so.
As discussed above, opacity arguments imply large values of the flow Lorentz factor $\Gamma$.  The question then arises as
to why $\Gamma$ is much larger than the Lorentz factors associated with patterns of radio emission.   One possibility, discussed 
in some detail below, is that dissipation of bulk energy results from collision of the fast fluid with the surrounding matter,
in quasi-stationary structures.
An alternative scenario \cite{lev07} is that flares observed in sources like Mrk 421, Mrk 501 and PKS 2155-304 
are produced by radiative deceleration of fluid shells expelled during violent ejection episodes.
The basic picture is that fluid shells accelerate to a Lorentz factor $\Gamma_0>>1$ at some radius $r_d\sim10^2-10^3 r_g$, at which dissipation of their bulk energy occurs.
The dissipation may be accomplished through formation of internal shocks in a hydrodynamic jet or dissipation of magnetic energy in a Poynting flux dominated jet \cite{rom92,lev98}.
The shocks may also result from a focusing of the outflow, as already suggested above, which can in principle 
give rise to dissipation in a nozzle with a very small cross-sectional radius. 

A basic question is whether the radiation field required to provide the drag that decelerate the flow can still be transparent enough
to allow the TeV photons produced in the process to escape the system.  This problem has been addressed in Ref. \cite{lev07}.  
The main conclusion is that for a reasonably flat energy distribution of nonthermal electrons, $dn_e/d\epsilon\propto \epsilon^{-q}$ with $q\le2$, extension of the distribution to a maximum energy $\epsilon_{\rm max}$ at which the pair production optical depth, $\tau_{\gamma\gamma}(\Gamma_0\epsilon_{\rm max})$, is a few is already sufficient to cause appreciable deceleration of the front.  In this case the major fraction of the dissipated energy is released in the VHE band, and so there is no missing energy issue.  The model is also consistent with the minimum jet power estimated from the resolved radio synchrotron emission on VLBI scales \cite{lev96}, which in case of Mrk 421 and Mrk 501 was found to be much smaller than the TeV luminosity.  As shown in Ref.~\cite{lev07} for the TeV blazars a background luminosity of $L_s\sim 10^{41}-10^{42}$ erg s$^{-1}$, roughly the luminosity of LLAGN, would lead to a substantial deceleration of the front 
and still be transparent enough to allow the TeV $\gamma$-rays produced by Compton scattering of the background photons
to escape to infinity.  
The ambient radiation field is most likely associated with the nuclear continuum source.  Propagation 
of the $\gamma$-ray flare from low-to-high energies, as reported recently for Mrk 501~\cite{alb07}, are naturally expected in this
model, since the $\gamma$-spheric radius increases quite generally with increasing $\gamma$-ray energy~\cite{bln95}.
The bulk Lorentz factor of the jet during states of low activity may be appreciably smaller than that of fronts expelled during violent ejection episodes.

\subsection{Intermediate scales}
A different possibility is that the TeV emission is produced at relatively large radii.  The size scale of the emission zone implied by rapid variability should then be much smaller than the distance from the black hole.  Since only a fraction $(d/a)^2$, where $d$ is the size of the emission region and $a$ is the cross-sectional radius of the jet, can be dissipated, it seems difficult to account for the characteristic luminosities observed during flares in TeV blazars, unless there is a way to channel a sizable fraction of the bulk energy into a small area.  A particular motivation to consider such a scenario comes from observations of the HST-1 knot in M87, a stationary radio feature associated with the sub-kpc scale jet.  The knot 
is located at a projected distance of 60 pc ($0.86''$) from the central engine, and is known to be 
a region of violent activity.   Sub-features moving away from the main knot of the HST-1 complex 
at superluminal speeds have been detected recently \cite{cheu07}.  In addition, rapid, large amplitude variations of 
the resolved X-ray emission from HST-1 have been reported, with doubling time $t_{\rm var}$ as short as $0.14$ yrs.
The observed variability limits the linear size of the X-ray source to
$\Delta r \simlt \Gamma D t_{\rm var}\sim 0.022\Gamma D$ pc, which for reasonable estimates of the 
Doppler factor is three orders of magnitude smaller than the distance between the HST-1 knot and the central black hole.
Based on a claimed correlation between the X-ray and TeV emission it has been proposed that HST-1 may also be the 
region where the TeV emission is produced \cite{cheu07}.  As mentioned above, this motivated the consideration that the TeV emission zone in TeV blazars may also be located far from the black hole.

It has been proposed that HST-1 reflects the location of a recollimation nozzle \cite{staw06,Lev08}.  
In this picture the superluminal sub-knots that seem to be expelled from the HST-1 complex can be associated with
internal shocks produced by reflection of the recollimation shock at the nozzle.  As stated above, the rapid variability sets a limit
on the cross-sectional radius of the channel at the location of HST-1, $a/r_{\rm HST1}\simlt 10^{-3}$  \cite{Lev08}.
As demonstrated in Ref.~\cite{Lev08} modest radiative cooling behind the recollimation shock can lead to 
a focusing of the relativistic jet, so that at the nozzle it has a very small cross-sectional radius.
This mechanism is most effective when the bulk energy of the inner, relativistic jet at the recollimation region 
is dominated by rest mass energy, and may account for some of the features observed in the TeV blazars.  However, the low 
radiative efficiency of the jet in M87 implies that this mechanism may not apply to this particular source.  Nonetheless, if
at a distance of several parsecs from the central engine the jet in M87 encounters a gaseous condensation having a 
flat pressure profile, as discussed in Ref.~\cite{staw06}, then it will remain well collimated and can have a sufficiently
small cross-sectional radius at the HST-1 location to account for the short variability time observed (Bromberg \& Levinson,
in preparation).

\acknowledgments
I thank C. Dermer and D. Eichler for discussions

\end{document}